# Cavity-tunable Pseudospin-Hall Topology and Edge Transport

Minghui Niu[1,#], Yan Meng[2,#,*], Chengle Fan[3], Xiaoyuan Jiao[3], Jidong Song[1], Perry Ping Shum[3], Zhen Gao[3,*], and Mengyao Li[1,*]

[1] Institute of Materials Research, Tsinghua University Shenzhen International Graduate School, Shenzhen, 518055, China.

[2]Marine Science and Technology Domain, Beijing Institute of Technology, Zhuhai 519088, China.

[3]State Key Laboratory of Optical Fiber and Cable Manufacturing Technology, Department of Electronic and Electrical Engineering, Guangdong Key Laboratory of Integrated Optoelectronics Intellisense, Southern University of Science and Technology, Shenzhen 518055, China

[#]These authors contributed equally to this work.

*Corresponding author. Email: meng_yan@bitzh.edu.cn (Y.M.); gaoz@sustech.edu.cn (Z.G.) ; mengyaoli@sz.tsinghua.edu.cn (M.L.)

## Abstract

Topological edge states are a hallmark of topological phases, with their existence tied to bulk band topology through bulk–edge correspondence. In photonic pseudospin-Hall systems, these states exhibit pseudospin–momentum locking, with lattice interactions and interface geometry determining each pseudospin's propagation direction. Reversing the pseudospin-dependent propagation direction without altering the lattice configuration or excitation handedness remains a fundamental challenge in controlling topological edge transport. Here, we theoretically and experimentally demonstrate cavity-tunable topological edge transport in a microwave metasurface embedded within a metallic cavity. Varying the cavity height reshapes resonator interactions, driving band inversion and topological phase transition while preserving the lattice geometry. For a fixed domain-wall configuration, this transition reverses the propagation direction associated with each pseudospin while preserving pseudospin–momentum locking. Excitations with the same handedness are observed to launch edge waves propagating in opposite directions at their respective edge-band frequencies determined by the cavity height. These results establish cavity electrodynamics as an external means of controlling the bulk topology and reconfiguring the directional response of topological interfaces, creating new opportunities for topological wave routing in fixed photonic lattice structures.

## Introduction

Topological edge transport draws its robustness from a distinction between bulk topological phases that persists while the band gap and protecting symmetries are maintained. Bulk–edge correspondence connects this distinction to the existence and connectivity of boundary modes[1–4]. Their directional transport and resilience to perturbations have motivated applications in waveguides[5-7], cavities[8], and lasers[9]. In photonic analogs of the quantum spin Hall effect, helical edge states associate opposite pseudospins with counterpropagating channels, enabling handedness-selective excitation[10–12]. Researchers typically establish the underlying band topology and pseudospin-to-direction correspondence through lattice design. Reconfigurable implementations commonly modify the structure or its constituent material properties to alter the interactions that determine the bands[13,14]. Switching the excitation handedness can also redirect edge waves by selecting the counterpropagating channel[15]. Reversing the propagation direction associated with a given excitation handedness while retaining the same lattice geometry[16], therefore, remains an important challenge in topological photonics.

External control of orbital band ordering offers a possible way to reconfigure helical transport while preserving the lattice and excitation handedness, motivating investigation of the connection between tunable bulk topology and the directional response of fixed interfaces. For instance, the electromagnetic environment can alter band topology without changing the lattice[17-21]. Theory predicts such a strategy may induce changes in Dirac dispersions[18], gauge fields[19], and topological transitions[20-22]; a one-dimensional transition has been experimentally observed[23]. However, how a cavity-induced and tunable electromagnetic environment governs mode helicity and pseudospin-dependent topological transport remains insufficiently understood. Specifically, whether cavity tuning can induce a reversal of the pseudospin-Hall response in the bulk and invert the pseudospin-dependent propagation of interface states without reconfiguring the lattice or domain wall remains an open question.

Here, we theoretically and experimentally demonstrate cavity-controlled orbital band inversion and topological helical edge transport in a breathing-honeycomb microwave metasurface keeping the in-plane lattice fixed. Cavity confinement redistributes the electrostatic and photon-mediated contributions to the resonator interactions, closing and reopening the bulk gap with inverted orbital ordering. Theory and full-wave calculations identify different renormalizations of the Γ-point $p$- and $d$-like doublets, driving them through an accidental double-Dirac crossing and reversing their ordering. Calculated topological indices characterize the resulting change in band topology, while momentum-resolved near-field spectroscopy resolves the accompanying bulk gap closing and reopening. Along an unchanged domain wall, a source with fixed position and handedness launches edge waves in opposite directions at the respective edge-band frequencies of the two gapped regimes. Pseudospin-resolved eigenmode calculations connect this response to an exchange of the propagation directions associated with the two helical sectors, with pseudospin–momentum locking retained. These results establish an experimental connection between cavity-driven orbital inversion and the reconfiguration of a fixed helical interface, providing a basis for tunable topological waveguides and pseudospin-selective routing.

## Results

We consider a honeycomb metasurface comprising a triangular array of hexagonal clusters, each containing six electric-dipole emitters/antennas[4,24]. The metasurface is confined between two conducting plates separated by a distance $L$ (Fig. 1**a**). In the unperturbed honeycomb limit, the two inequivalent valleys fold onto the Γ point of the reduced Brillouin zone associated with the six-site unit cell, producing a fourfold double-Dirac degeneracy[5]. The corresponding modes form $p$- and $d$-like doublets, whose chiral combinations define two photonic pseudospin sectors[24,25]. We denote the distance from each resonator to its cluster center by $R$, with $R_0$ representing its value in the unperturbed lattice. The displacement $\delta R = R - R_0$ is negative for a shrunken lattice and positive for an expanded lattice. This breathing deformation modifies the balance between intra- and intercluster coupling, lifting the degeneracy between the two doublets while preserving sixfold rotational symmetry. The in-plane resonator positions remain fixed as $L$ is varied symmetrically about the metasurface plane.

The dipole–dipole interactions nevertheless vary substantially with cavity height. In a wide cavity, the subwavelength dipole spacing makes electrostatic coupling ($\mathcal{V}^{\mathrm{es}}$) the dominant contribution to the collective bands near the Dirac point. As the cavity narrows, the longitudinal electrostatic coupling between nearby dipoles (red line) is progressively suppressed, while that between distant pairs (grey line) remains weak (Fig. 1**b**). Photon-mediated radiative coupling ($\mathcal{V}^{\mathrm{ph}}$) exhibits a distinct dependence on dipole separation: with decreasing cavity height, it grows steadily for distant pairs (grey line) but reaches a maximum before decreasing for nearby pairs (red line), as shown in Fig. 1**c**. Cavity tuning therefore modifies both the balance between electrostatic and photon-mediated radiative interactions and their spatial dependence, rather than simply rescaling all couplings uniformly. This redistribution of interaction strengths and ranges reshapes the collective bands without altering the lattice geometry[26–28].

In our metasurface, we describe the interactions in the six-site Bloch basis with a coupling matrix[22],

$$\mathcal{M}_{ij}(k, \omega, L) = \sum_{R}{}' \mathcal{V}\,(\rho_{ij}, R, \omega L)\ \exp\ \left[-ik \cdot \rho_{ij}(R)\right] \tag{1}$$

where $R$ is a Bravais-lattice vector identifying a unit cell, and the basis positions $d_i$ identify the electric dipoles within it. The coupling matrix sums the interactions between all distinct electric dipoles across the lattice, including those within the same unit cell, while excluding each dipole's self-interaction. We decompose the interaction kernel in Eq. (1) into electrostatic and photon-mediated radiative contributions, $\mathcal{V} = \mathcal{V}^{\mathrm{es}} + \mathcal{V}^{\mathrm{ph}}$. The collective band structure is determined by the poles of the cavity-modified coupled-dipole response[22]:

$$\det[\chi_{\mathrm{cav}}^{-1}(\omega, L)\mathbb{I}_6 - \mathcal{M}(k, \omega, L)] = 0 \tag{2}$$

with $\chi_{\mathrm{cav}}^{-1}(\omega, L) = \chi_0^{-1}(\omega) - \Sigma_{\mathrm{self}}\,(\omega, L)$ . Here, $\Sigma_{\mathrm{self}}$ accounts for the cavity-induced renormalization of an isolated dipole, while $\mathcal{M}$ describes the collective coupling between distinct sites. Solving Eq. (2) self-consistently for $\omega$ at each Bloch wavevector incorporates both the local cavity dressing and the frequency-dependent intersite coupling within the same eigenvalue problem (see details in Methods).

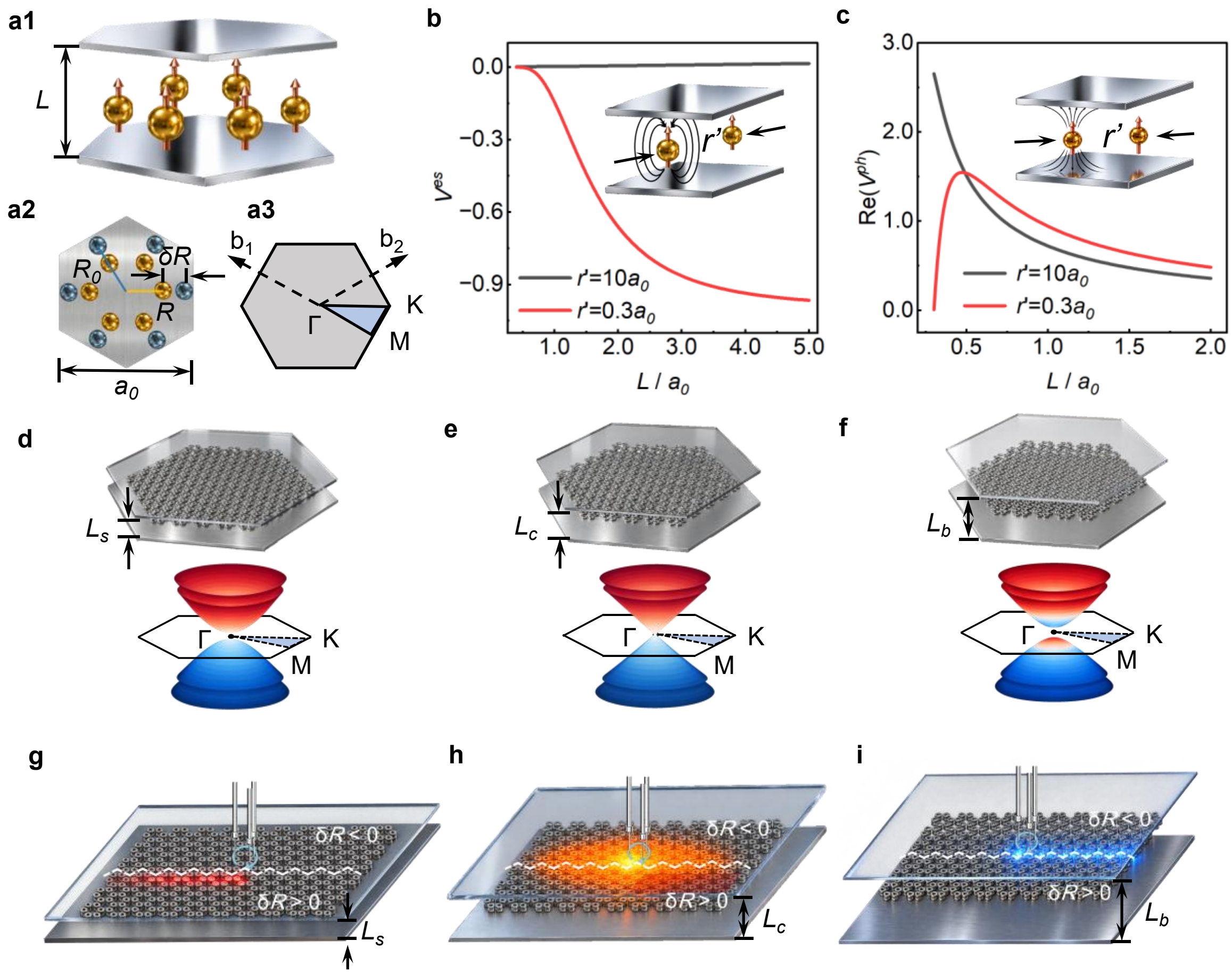

**Fig. 1 | Cavity-tunable dipole-dipole interaction, orbital band inversion, and helical edge transport. a1,** Schematic of the unit cell of a breathing-honeycomb metasurface composed of interacting dipole emitters/antennas embedded inside a metallic cavity with a height $L$. **a2**, Six-site artificial atom. **a3**, Corresponding first Brillouin zone. **b**,**c**, Cavity-induced redistribution of the dipole–dipole interaction of the electrostatic coupling (**b**) and photon-mediated radiative coupling (**c**) for $r' = 10a_0$ (grey line) and $0.3a_0$ (red line) versus cavity height $L/a_0$. **d**–**f**, Evolution of the bulk bands in a fixed shrunken lattice, from a gapped phase (**d**) through an accidental double-Dirac point (**e**) to a reopened gap with inverted $p-d$ ordering (**f**). **g**–**i**, Evolution of helical edge transport along the interface between shrunken and expanded domains under excitation by the same chiral source. The spectrally isolated edge channel is lost at bulk gap closure (**h**) and re-emerges in the reopened gap with reversed pseudospin-to-direction correspondence (**i**).

Figures 1**d**–**f** schematically illustrate the evolution of the bulk bands with cavity height at fixed lattice geometry. The $p$- and $d$-like orbital doublets become degenerate at a critical plate separation (Fig. 1**e**), forming an accidental double-Dirac point, and exchange order as the gap reopens (Fig. 1**f**). The bands

calculated from Eq. (2) reproduce this sequence (Figs. S1**a**–**c**), supporting cavity-driven orbital band inversion. This bulk-band evolution also governs the interface response illustrated in Figs. 1**g**–**i**. A fixed domain wall between shrunken ($\delta R < 0$) and expanded ($\delta R > 0$) lattices (white curved line) supports a pair of counterpropagating helical edge states when the two domains have distinct band topology[29–31]. At the transition, closure of the common bulk gap removes the spectral isolation of the edge channel and the helical edge states become bulk states (Fig. 1**h**). The channel re-emerges in the reopened gap with reversed pseudospin-to-direction correspondence (Fig. 1**i**). Pseudospin–momentum locking is retained in both gapped regimes, while a chiral source with unchanged handedness selects opposite propagation directions at the respective edge-band frequencies. This predicted sequence connects cavity-driven orbital inversion to the loss and recovery of a spectrally isolated edge channel and the reversal of directional transport along a fixed interface.

We next implement this mechanism in an experimentally accessible metasurface comprising cap-loaded microwave helical resonators arranged in a breathing-honeycomb lattice between two parallel conducting plates (Figs. 2**a**–**c**). Each resonator acts as an effective electric dipole. The lattice period is $a_0 = 48$ mm, and the plate separation $L$ defines the cavity height. The resonator geometry is specified by $t$, $r$, $R_1$, $d$, $h$ and $R_2$, as detailed in Methods. Its dominant mode is polarized normal to the metasurface, with the electric field concentrated in the gap between the cap and the cavity plate, realizing the out-of-plane dipolar response assumed in the analytical model[19,22]. Crucially, $L$ can be varied independently of the in-plane resonator arrangement.

For the fixed shrunken lattice, the full-wave eigenmode calculations (dotted line) agree well with the analytical bands (solid line) at three cavity heights, as shown in Figs. 2**d**–**f**. The agreement both at Γ and away from this high-symmetry point indicates that the reduced model captures the momentum-dependent collective electrodynamics of the physical resonator array. Under strong confinement with a small cavity height $L_s$, the $p$- and $d$-like doublets are separated by a finite gap at Γ (Fig. 2**d**). Increasing the cavity height renormalizes the two orbital sectors differently until they become degenerate at $L_c$ (Fig. 2**e**), producing a fourfold crossing with approximately linear dispersion near Γ. Because the lattice retains its breathing deformation, this crossing is an accidental degeneracy induced by electrodynamic tuning[22], distinct from the symmetry-enforced, zone-folded double-Dirac point of the unperturbed honeycomb lattice[4]. Further increasing $L$ beyond $L_c$ reopens the gap with the $p$- and $d$-like sectors interchanged (Fig. 2**f**), yielding orbital band inversion without structural deformation.

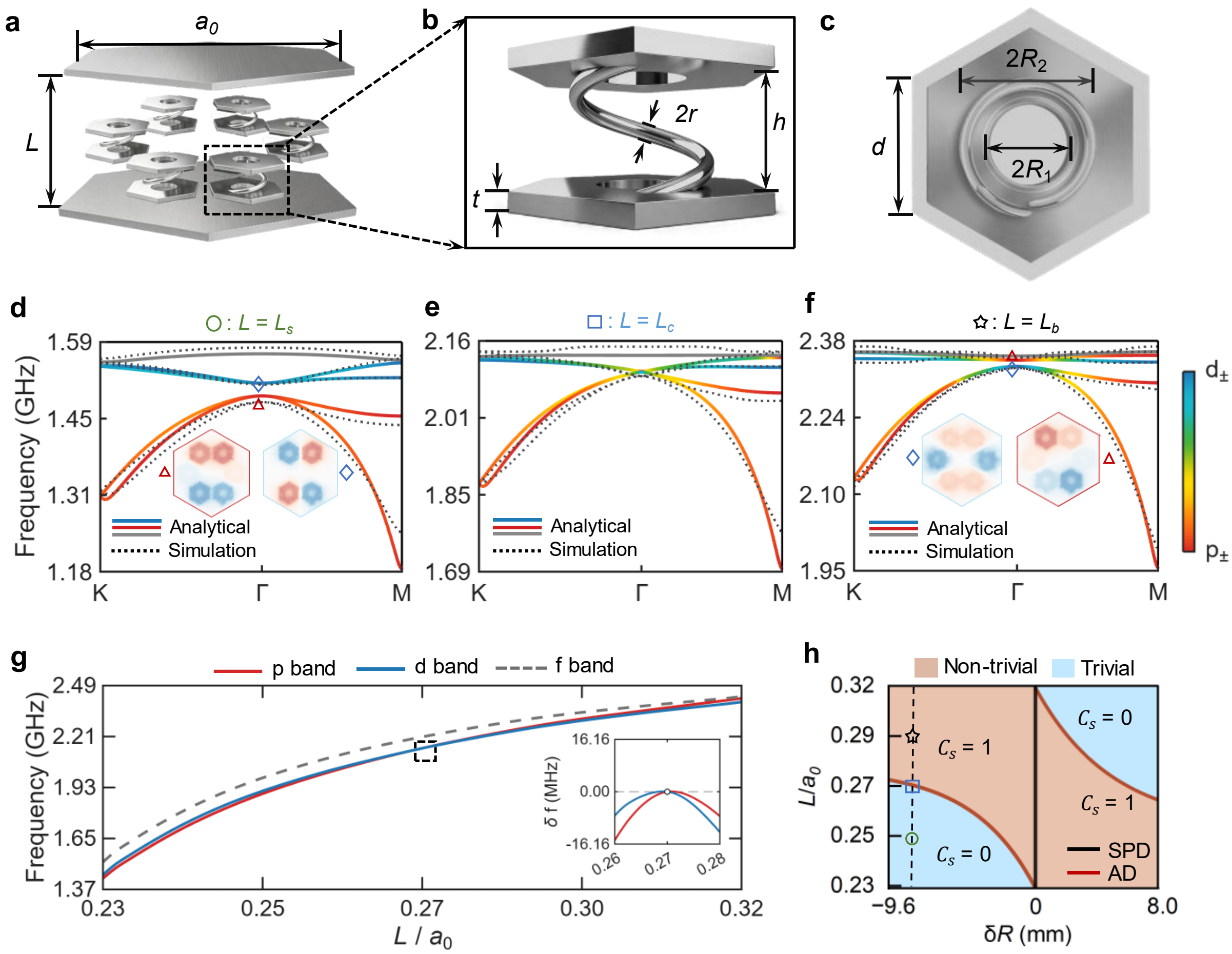

**Fig. 2 | Cavity-controlled orbital band inversion and topological phase transition. a**, Breathing-honeycomb metasurface comprising cap-loaded microwave resonators (CLMRs) confined within a tunable metallic cavity. **b**,**c**, Geometry and structural parameters of an individual CLMR. **d**–**f**, Comparison of the analytical (solid line) and simulated (dotted line) band structures for $L_s/a_0 = 0.25$, $L_c/a_0 = 0.27$ and $L_b/a_0 = 0.29$, respectively, at fixed $\delta R = -0.9$ mm. The gap closes at $L_c$ and reopens with reversed $p-d$ ordering. **g**, Γ-point frequencies of the $p$-, $d$- and $f$-like bands as a function of cavity height $L$. **h**, Topological phase diagram in the $(\delta R, L/a_0)$ parameter space, distinguishing symmetry-enforced and cavity-induced topological phase transition.

The red and black lines denote the symmetry-protected fourfold degeneracy (SPD) and accidental fourfold degeneracy (AD), respectively.

The orbital selectivity of the cavity-induced renormalization is evident in Fig. 2**g**, which tracks the Γ-point frequencies of the $p$- (red line), $d$- (blue line), and $f$-like (grey line) bands as a function of cavity height $L$. The $p$- and $d$-like doublets shift at different rates, become degenerate at $L_c$, and exchange their ordering, while the $f$-like bands remain spectrally separated from them. This topological phase transition is therefore governed predominantly by the low-energy $p - d$ subspace. Projecting the full electrodynamic problem onto this subspace[32,33] yields an effective Dirac Hamiltonian whose mass comprises electrostatic and photon-mediated radiative contributions, $M_{es}$ and $M_{ph}$, respectively, as detailed in Supplementary Notes 3-6:

$$M_{\text{total}}(L) = M_{ph}(L) + M_{es}(L). \tag{3}$$

These contributions respond differently to cavity confinement, causing the effective Dirac mass to change sign at the critical cavity height $L_c$[22], defined by $M_{\text{total}}(L_c) = 0$ (Fig. S2**a**). For the fixed shrunken lattice, the spin-Chern number calculated from the effective Hamiltonian changes from 0 to 1 across this transition, identifying the two gapped regimes as topologically distinct[4] (Fig. S2**b**). The evolution of the Γ-point $p - d$ splitting thus provides a direct spectral signature of orbital band inversion, while the calculated spin-Chern number characterizes the accompanying cavity-controlled topological phase transition.

The phase diagram shows how lattice deformation ($\delta R$) and cavity height ($L$) jointly determine the bulk topology (Fig. 2**h**). The gapped regions are characterized by bulk topological invariants $C_s = 0$ (blue) and $C_s = 1$ (orange) (see details in Supplementary Note 7 and Figs. S3**a-b**). The black line at $\delta R = 0$ marks the symmetry-protected double-Dirac condition produced by Brillouin zone folding. The red boundaries at finite lattice deformation denote accidental double-Dirac conditions, where cavity tuning brings the $p$- and $d$-like doublets into degeneracy and closes the bulk band gap. The dashed vertical line crosses one of these boundaries, indicating that changing only the cavity height can switch the band topology of the same shrunken lattice.

We next experimentally probe the bulk-band evolution associated with the cavity-tunable topological phase transition. In the experimental setup shown in Fig. 3**a**, a finite breathing-honeycomb metasurface is embedded between two parallel metallic plates. The cavity height is varied by adjusting the plate separation while keeping the metasurface geometry fixed. An electric monopole antenna provides local excitation, while a probe is scanned across the resonator sites to measure the complex out-of-plane electric field $E_z$. The rhombic sample contains 11 × 11 unit cells, each comprising six cap-loaded resonators (Fig. 3**b**). At each excitation frequency, we record the complex field at the resonator sites and Fourier transform the measured electric field distributions to reconstruct the bulk band dispersion along the $\mathrm{K} - \Gamma - \mathrm{M}$ path in the two-dimensional Brillouin zone[34]. Figures 3**c**–**e** present the measured (color maps) and simulated (green dotted line) bulk band structures at three different cavity heights. We can observe that, as we increase the cavity height, the photonic band gap first closes (Fig. 3**d**) and then reopens (Fig. 3**e**). The accompanying inversion of the $p$- and $d$-like orbital ordering is identified from the mode-resolved analytical and full-wave calculations.

To further examine the experimental signatures the cavity-driven topological phase transition, we reconstruct the Wannier bands with the aid of the calculated Bloch eigenmodes. For each of the six site positions within a unit cell, we Fourier transform the fields measured at the corresponding sites across the array. At each momentum and frequency, the resulting six complex amplitudes retain their relative phases and define a six-component field vector[35]. We project this vector onto the calculated eigenmodes below the gap. For each fixed $k_1$, we evaluate a Wilson loop over the transverse reciprocal-space coordinate[36,37]. The resulting eigenphases are then plotted as a function of $k_1$, defining the hybrid Wannier bands[36,38]. Applying the same procedure to calculated and projected experimental states gives a model-assisted representation of the band-subspace connectivity.

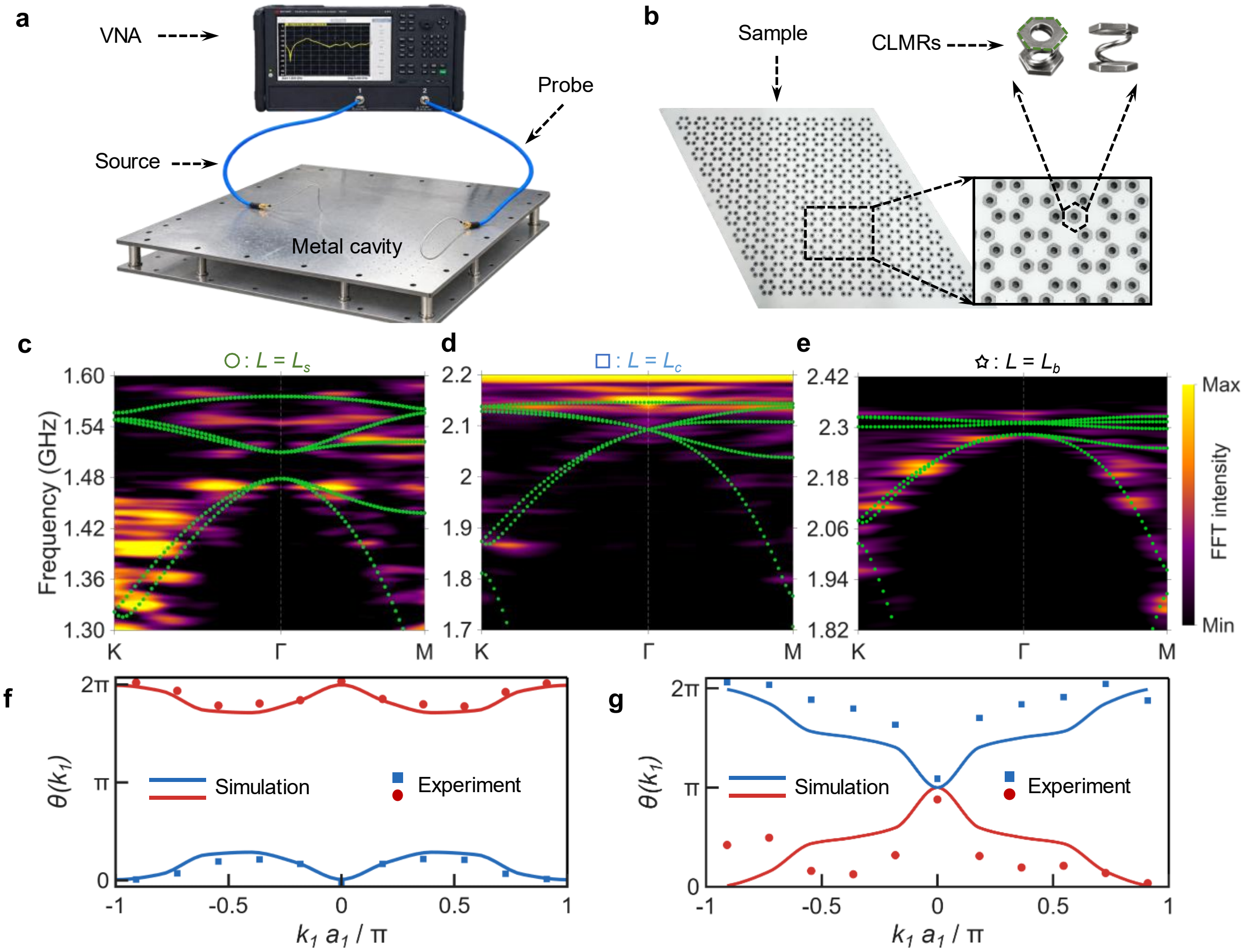

**Fig. 3 | Experimental observation of the cavity-controlled topological phase transition. a**, Near-field measurement setup with a vector network analyser, a fixed source and a movable electric-field probe. **b**, Rhombic sample comprising 11 × 11 breathing-honeycomb unit cells and the detail of CLMRs . **c**–**e**, Measured (colormaps) and simulated (green dotted line) momentum-resolved bulk band structures at cavity heights $L_\mathrm{s}$ , $L_\mathrm{c}$ and $L_\mathrm{b}$, respectively. **f**,**g**, Measured (red dots and blue squares) and calculated (red and blue solid lines) Wannier bands at $L_\mathrm{s}$ and $L_\mathrm{b}$, respectively, exhibiting the transition from trivial Wannier-centre without partner switching to nontrivial flow with partner switching.

For $L = L_\mathrm{s}$ (Fig. 3**f**), the calculated Wannier centres (red and blue solid lines) and their model-assisted experimental reconstructions (red dots and blue squares) remain separated and do not exchange partners across the Brillouin zone, consistent with the trivial phase predicted by the effective theory. For $L = L_\mathrm{b}$ (Fig. 3**g**), the reconstructed Wannier branches approach $\theta = \pi$ and exchange partners near $k_1 = 0$, reproducing the connectivity of the calculated Wilson-loop spectrum (see details in Supplementary Fig. S3**a-b**). Together with the directly observed closing and reopening of the bulk gap, these results connect the measured band evolution to the theoretically identified topological transition. Thus, the accompanying reversal of topological phase under unchanged lattice configuration is demonstrated experimentally.

Finally, we experimentally demonstrate cavity-tunable helical edge transport. Bulk–edge correspondence[1,38] connects changes in bulk topology to a reorganization of the interface spectrum. To probe the directional consequences of this reorganization, we construct a fixed domain wall between shrunken and expanded lattices and vary the cavity height, testing whether the electromagnetic environment can reverse helical edge propagation without changing the excitation handedness. The sample comprises $11 \times 6$ breathing-honeycomb unit cells (Fig. 4**a**). Shrunken and expanded domains with $\delta R = -0.9$ mm and 0.8 mm, respectively, form a zigzag domain wall (white dashed line) through the centre of the sample. Figure 4**b** shows the two unit-cell geometries and their interface. Three electric monopole antennas at fixed positions adjacent to the domain wall synthesize a rotating near-field excitation, with successive phase offsets of $2\pi/3$. The phase sequence $0, 2\pi/3, 4\pi/3$, or its reverse, produces clockwise or anticlockwise excitation, respectively. A probe scans the resonator sites throughout the sample to record the complex out-of-plane electric field $E_z$.

In the short-cavity configuration, $L = L_\mathrm{s}$, clockwise chiral excitation predominantly launches the left-moving interface channel at 1.45 GHz. The simulated and measured electric fields remain localized along the zigzag domain wall and show leftward propagation (Fig. 4**c**,**d**). Fourier transformation of the measured complex electric fields along the interface yields the momentum-resolved spectrum in Fig. 4**e**, whose dominant spectral weight follows the calculated branch with negative group velocity. The real-space field maps and momentum-resolved spectrum thus consistently identify predominantly leftward edge transport. Increasing the cavity height to $L = L_\mathrm{b}$ reverses this directional response. With the chiral source and domain-wall geometry unchanged, the simulated and measured electric fields show predominantly rightward propagation at 2.27 GHz (Fig. 4**f**,**g**). The dominant spectral weight correspondingly shifts to the interface branch with positive group velocity (Fig. 4**h**). Reversing the excitation handedness selects the counterpropagating channel within each cavity configuration: anticlockwise chiral excitation launches a right-moving edge mode in both simulation and experiment at $L = L_s$ (Supplementary Fig. S4**a**,**b**) and a left-moving mode at $L = L_b$ (Supplementary Fig. S4**c**,**d**). Together with the pseudospin-resolved eigenmode simulations in Supplementary Fig. S5**a**,**b**, these results support a cavity-induced reversal of the pseudospin-to-direction correspondence. Pseudospin–momentum locking is retained in both gapped regimes, while the propagation direction selected by a given excitation handedness reverses. The two gapped cavity configurations therefore provide opposite directional operating states of the same interface. Supplementary Fig. S6 additionally shows a delocalized bulk-field response at $L = L_c$, complementing the interface-localized fields in Fig. 4. The demonstrated reversal follows the cavity-induced reorganization of the topological edge-state bands and occurs at their respective edge-band frequencies, with the in-plane lattice, domain wall, and excitation handedness unchanged.

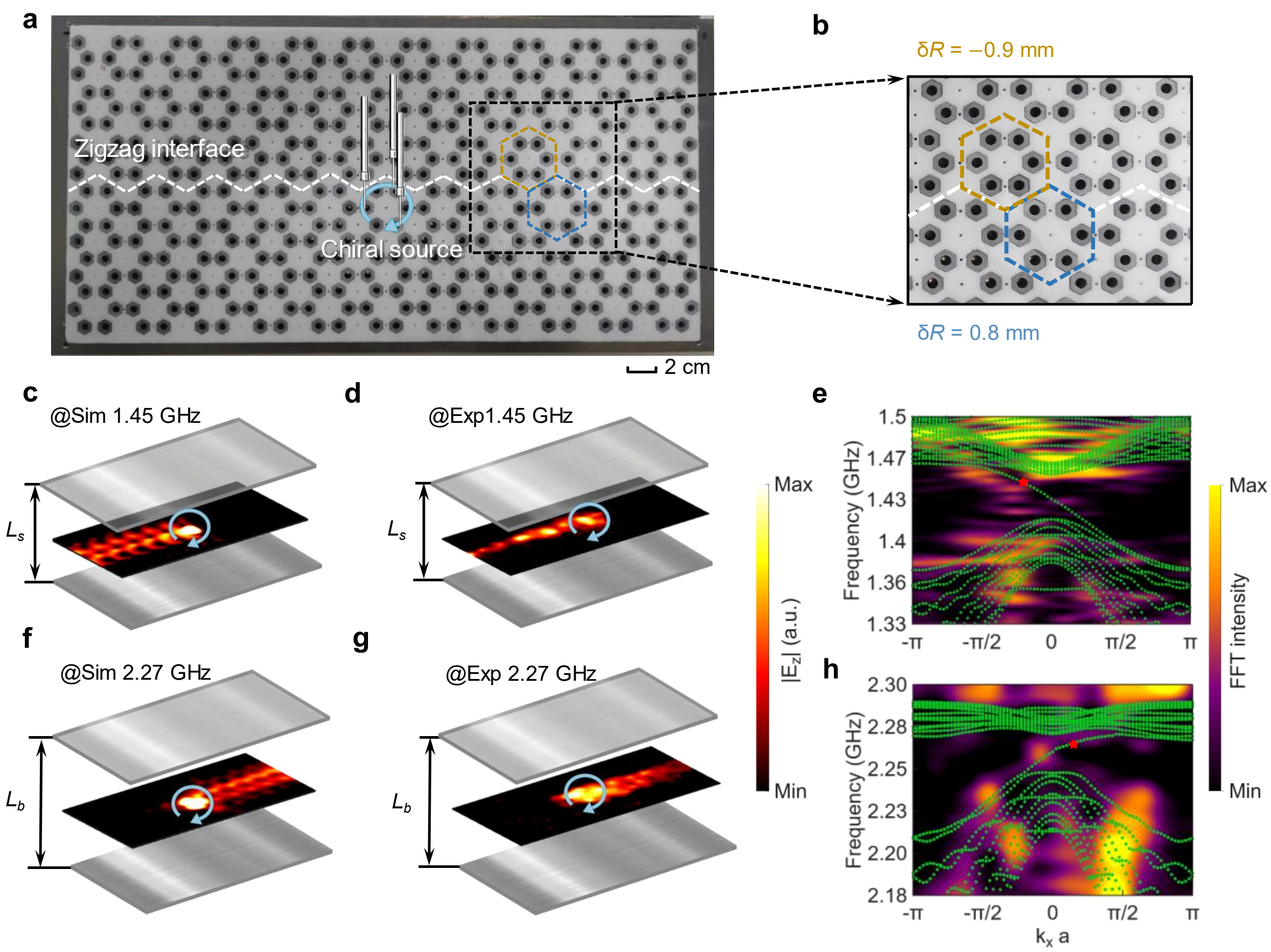

**Fig. 4 | Experimental demonstration of cavity-tunable helical edge transport.** **a**, Photograph of the sample with a zigzag interface (white dashed line) between shrunken ($\delta R = -0.9$ mm) and expanded ($\delta R = 0.8$ mm) domains. **b**, Enlarged view of the zigzag interface. **c**,**d**, Simulated and measured electric-field distributions at $L = L_\mathrm{s}$ and $f$ = 1.45 GHz under clockwise chiral excitation, showing predominantly leftward helical edge propagation. **e**, Corresponding measured momentum-resolved interface spectrum (colour map) overlaid with simulated bands (green dotted curves), showing dominant spectral weight on the branch with negative group velocity. **f**,**g**, Simulated and measured electric-field distributions at $L = L_\mathrm{b}$ and $f$ = 2.27 GHz, with the source position and clockwise excitation handedness unchanged, showing predominantly rightward helical edge propagation. **h**, Corresponding measured interface spectrum (colour map) overlaid with simulated bands (green dotted curves), showing the transfer of dominant spectral weight to the branch with positive group velocity.

## Discussion

We have theoretically and experimentally demonstrated cavity-controlled orbital band inversion and helical edge transport in a fixed breathing-honeycomb microwave metasurface. Varying the cavity height reshapes the relative strengths and spatial ranges of electrostatic and photon-mediated interactions, causing the Γ-centred $p$- and $d$-like doublets to become degenerate and exchange their ordering without changing the in-plane lattice. Analytical and full-wave calculations identify an accidental double-Dirac crossing and the accompanying topological phase transition, while momentum-resolved near-field spectroscopy reveals the bulk gap closing and reopening in the same fabricated sample. The band topology is characterized using calculated eigenstates and Wannier bands reconstructed from measured electric fields.

Our results extend environmental control to a distinct orbital pseudospin platform, connecting Γ-centred $p - d$ inversion to the directional response of a helical edge interface. Both gapped regimes retain pseudospin–momentum locking, while the propagation direction associated with each pseudospin reverses. Consequently, a chiral source with unchanged position and handedness launches predominantly opposite propagation directions along the same domain wall at the respective edge-band frequencies. Together, the bulk spectroscopy, directional transport measurements, and pseudospin-resolved calculations establish a connection between cavity-driven orbital inversion and cavity-tunable helical edge transport. The calculations further show that bulk gap closure removes the spectral isolation of the interface channel, linking environmental tuning to both its directional response and its separation from bulk modes.

These findings establish the electromagnetic environment as a design parameter for orbital band topology and pseudospin-selective transport in an existing resonator lattice. Cavity control complements programmable and optically tunable photonic platforms[41–43] by enabling reconfiguration through electromagnetic boundary conditions. Future spatially patterned or time-dependent cavity configurations could extend this approach towards locally controlled topological interfaces and reconfigurable wave-routing networks[44].

**Acknowledgements**

M.L. thanks financial support from the National Key R&D Program of China (Grant No. 2023YFB3811400), National Natural Science Foundation of China (Grant No. 12404435), Guangdong Science and Technology Commission (Grant No.2025A1515010774). Z.G. acknowledges funding from the National Key R&D Program of China under Grant No. 2025YFA1412300, National Natural Science Foundation of China under Grant No. 62361166627 and 62375118, Guangdong Basic and Applied Basic Research Foundation under Grant No. 2024A1515012770, Shenzhen Science and Technology Innovation Commission under Grants no. 20230802205352003, and High-level Special Funds under Grant No. G03034K004. Y.M. acknowledges the support from the National Natural Science Foundation of China under Grant No. 12674380 and 12304484, Guangdong Basic and Applied Basic Research Foundation under Grant No. 2024A1515011371.

## Data Availability

The data that support the findings of this study are available from the corresponding author upon reasonable request.

## Competing interests

The authors declare no conflict of interest.